\documentclass[runningheads]{llncs}

\def\ket#1{\mathinner{|{#1}\rangle}}

\def\Ket#1{\left|#1\right>}
{\catcode`\|=\active 
  \gdef\Braket#1{\begingroup \mathcode`\|32768\let|\BraVert\left<{#1}\right>\endgroup}
}
\def\BraVert{\egroup\,\mid@vertical\,\bgroup}
{\catcode`\|=\active
  \gdef\set#1{\mathinner{\lbrace\,{\mathcode`\|"8000\let|\midvert #1}\,\rbrace}}
  \gdef\Set#1{\left\{\:{\mathcode`\|"8000\let|\SetVert #1}\:\right\}}}
\def\midvert{\egroup\mid\bgroup}
\def\SetVert{\egroup\;\mid@vertical\;\bgroup}

\usepackage{graphicx}
\usepackage{amsfonts}
\usepackage{amsmath}
\usepackage{latexsym}
\usepackage{amssymb}
\usepackage[mathcal]{euscript}
\usepackage[table]{xcolor}

\newcommand\VD[1]{} 	
\newcommand\VND[1]{#1} 	
\newcommand\VL[1]{#1} 	
\newcommand\VS[1]{} 	

\newcommand{\Z}{\mathbb{Z}}

\newcommand{\pa}[1]{\left(#1\right)}

\newcommand{\eg}{\emph{e.g.\ }{}}
\newcommand{\ie}{\emph{i.e.\ }{}}
\newcommand{\cf}{\emph{cf.\ }{}}

\newcommand{\Phase}{R($\frac{\pi}{4})${}}
\newcommand{\cPhase}{controlled-\Phase{}}

\newcommand{\cells}[4]{
  \centering
  \Ket{
   \,
  \begin{tabular}{ | p{2.8mm} | p{2.8mm} | }
   \hline			
     #1 & #2   \\ \hline
     #3 & #4   \\ \hline
   \end{tabular}\,
  }
}

\newcommand{\barrier}{\cellcolor{orange}}

\newtheorem{Cl}{Claim}

\newtheorem{Def}{Definition}

\begin{document}

\title{A Simple $n$-Dimensional Intrinsically Universal Quantum Cellular Automaton}
\author{Pablo Arrighi \and Jonathan Grattage}

\institute{University of Grenoble, \VL{Laboratoire }LIG,\VL{\\}
 \VL{B\^{a}timent IMAG C, }220 rue de la Chimie,\VL{\\}
 38400 \VS{SMH}\VL{Saint-Martin-d'H\`eres}, France
 \and
 \VL{Ecole Normale Sup\'erieure de }\VS{ENS-}Lyon, \VL{Laboratoire }LIP,\VL{\\}
 46 all\'ee d'Italie, 69364 Lyon cedex 07, France}

\titlerunning{A Simple $n$-Dimensional Intrinsically Universal QCA}
\authorrunning{P. Arrighi and J. Grattage}

\toctitle{A Simple $n$-Dimensional Intrinsically Universal Quantum Cellular Automaton}
\tocauthor{Pablo Arrighi and Jonathan Grattage}

\maketitle

\begin{abstract}
We describe a simple $n$-dimensional quantum cellular automaton (QCA) capable of simulating all others,
in that the initial configuration and the forward evolution of any $n$-dimensional QCA can be encoded within the initial configuration of the intrinsically universal QCA. Several steps of the intrinsically universal QCA then correspond to one step of the simulated QCA. The simulation preserves the topology in the sense that each cell of the simulated QCA is encoded as a group of adjacent cells in the universal QCA.
\end{abstract}

\section{Introduction}\label{sec:introduction}\label{subsec:CA}

Cellular automata (CA), first introduced by Von Neumann \cite{Neumann}, consist of an array of identical cells, each of which may take one of a finite number of possible states. The whole array evolves in discrete time steps by iterating a function $G$. This global evolution $G$ is shift-invariant (it acts everywhere the same) and local (information cannot be transmitted faster than some fixed number of cells per time step).
Because this is a physics-like model of computation \cite{MargolusPhysics}, Feynman \cite{FeynmanQCA}, and later Margolus \cite{MargolusQCA}, suggested 
that quantising this model was important, for two reasons: firstly, because in CA computation occurs without extraneous (unnecessary) control, hence eliminating a source of decoherence; and secondly because they are a good framework in which to study the quantum simulation of a quantum system. From a computation perspective there are other reasons to study QCA,  such as studying space-sensitive problems in computer science, \eg `machine self-reproduction' \cite{Neumann} or `Firing Squad Synchronisation', which  QCA allow in the quantum setting.  There is also a theoretical physics perspective, where CA are used as toy models of quantum space-time \cite{LloydQG}. The first approach to defining QCA \cite{ArrighiMFCS,DurrWell,Watrous} was later superseded by a more axiomatic approach \cite{ArrighiUCAUSAL,ArrighiLATA,SchumacherWerner} together with the more operational approaches \cite{BrennenWilliams,NagajWocjan,PerezCheung,Raussendorf,VanDam,Watrous}.

The most well known CA is Conway's `Game of Life', a two-dimensional CA which has been shown to be universal for computation, in the sense that any Turing Machine (TM) can be encoded within its initial state and then executed by evolution of the CA. Because TM have long been regarded as the best definition of `what an algorithm is' in classical computer science, this result could have been perceived as providing a conclusion to the topic of CA universality. This was not the case, because CA do more than just running any algorithm. They run distributed algorithms in a distributed manner, model phenomena together with their spatial structure, and allow the use of the spatial parallelism inherent to the model. These features, modelled by CA and not by TM, are all interesting, and so the concept of universality must be revisited in this context to account for space. This is achieved by returning to the original meaning of the word \emph{universality} \cite{AlbertCulik,Banks,DurandRoka}, namely the ability for one instance of a computational model to be able to simulate other instances of the same computational model. Intrinsic universality formalises the ability of a CA to simulate another in a space-preserving manner \cite{MazoyerRapaport,OllingerJAC,Theyssier}, and was extended to the quantum setting in \cite{ArrighiUQCA,ArrighiNUQCA,ArrighiPQCA}.

There are several related results in the CA literature. For example, refs. \cite{MargolusPhysics,MoritaCompUniv1D,MoritaCompUniv2D} provide computation universal Reversible Partitioned CA constructions, whereas ref. \cite{MoritaIntrinsicUniv1D} deals with their ability to simulate any CA in the one-dimensional case. The problem of minimal intrinsically universal  CA  was addressed, \cf \cite{OllingerRichard}, and for Reversible CA (RCA) the issue was tackled by Durand-Lose \cite{Durand-LoseLATIN,Durand-LoseIntrinsic1D}. The difficulty is in having an $n$-dimensional RCA simulate all other $n$-dimensional RCA and not, say, the $(n-1)$-dimensional RCA, otherwise a history-keeping dimension could be used, as by Toffoli \cite{ToffoliConstruction}. There are also several other QCA related results. Watrous \cite{WatrousFOCS} has proved that QCA are universal in the sense of QTM. Shepherd, Franz and Werner \cite{ShepherdFranz} defined a class of QCA where the scattering unitary $U_i$ changes at each step $i$ (CCQCA). Universality in the circuit-sense has already been achieved by Van Dam \cite{VanDam}, Cirac and Vollbrecht \cite{VollbrechtCirac}, Nagaj and Wocjan \cite{NagajWocjan} and Raussendorf \cite{Raussendorf}. In the bounded-size configurations case, circuit universality coincides with intrinsic universality, as noted by Van Dam \cite{VanDam}. QCA intrinsic universality in the one-dimensional case is resolved in ref. \cite{ArrighiFI}. Given the crucial role of this in classical CA theory, the issue of intrinsic universality in the $n$-dimensional case began to be addressed in refs. \cite{ArrighiNUQCA,ArrighiPQCA}, where it was shown that a simple subclass of QCA, namely Partitioned QCA (PQCA), are intrinsically universal. Having shown that PQCA are intrinsically universal, it remains to be shown that there exists a $n$-dimensional PQCA capable of simulating all other $n$-dimensional PQCA for $n>1$, which is presented here.

PQCA are QCA of a particular form, where incoming information
is scattered by a fixed unitary $U$ before being redistributed. Hence the problem of finding an intrinsically universal PQCA reduces to finding some universal scattering unitary $U$ (this is made formal in section \ref{subsec:flat}, see Fig.\ref{fig:flattening34}). Clearly the universality requirement on $U$ is much more difficult than just quantum circuit universality. This is because the simulation of a QCA $H$ has to be done in a parallel, space-preserving manner. Moreover we must simulate not only an iteration of $H$ but several ($H^2$, \ldots), so after every simulation the universal PQCA must be ready for a further iteration.

From a computer architecture point of view, this problem can be recast in terms of finding some fundamental quantum processing unit which is capable of simulating any other network of quantum processing units, in a space-preserving manner. From a theoretical physics perspective, this amounts to specifying a scattering phenomenon that is capable of simulating any other, again in a space-preserving manner.


\VD{ 
\section{Definitions} \label{definitions}

\VL{\subsection{$n$-Dimensional (Partitioned) QCA}}\label{subsecdef}
First we recall the necessary definitions for $n$-dimensional QCA. 
Config\-urations hold the basic states of an  entire array of cells, and hence denote the possible basic states of the entire QCA:
\begin{Def}\textbf{(Finite configurations)}
A \emph{(finite) configuration} $c$ over $\Sigma$ is a function $c:
\Z^n \longrightarrow \Sigma$, with $(i_i,\ldots,i_n)\longmapsto
c(i_i,\ldots,i_n)=c_{i_i\ldots i_n}$, such that there exists a (possibly empty)
finite set $I$ satisfying $(i_i,\ldots,i_n)\notin I\Rightarrow c_{i_i\ldots i_n}=q$, where $q$ is a distinguished \emph{quiescent} state of $\Sigma$.
The set of all finite configurations over $\Sigma$ will be denoted $\mathcal{C}^{\Sigma}_f$.
\end{Def}
Since this work relates to QCA rather than CA, the global state of a QCA can be a superposition of these configurations.
To construct the separable Hilbert space of superpositions of configurations the set of configurations must be countable.
This is why finite, unbounded, configurations are considered. 
\vspace{-2mm}\begin{Def}\textbf{(Superpositions of configurations)}\label{superp}
Let $\mathcal{H}_{\mathcal{C}^{\Sigma}_f}$ be the Hilbert space of configurations. Each finite configuration $c$ is associated with a unit vector $\ket{c}$, such that the family $\pa{\ket{c}}_{c\in\mathcal{C}^{\Sigma}_f}$ is an orthonormal basis of $\mathcal{H}_{\mathcal{C}^{\Sigma}_f}$. A \emph{superposition of
configurations} is then a unit vector in $\mathcal{H}_{\mathcal{C}^{\Sigma}_f}$.
\end{Def}
Detailed explanation of these definitions, as well as axiomatic definitions of QCA, are available \cite{ArrighiUCAUSAL,ArrighiLATA,SchumacherWerner}. Building upon these works, we have shown \cite{ArrighiNUQCA,ArrighiPQCA} that Partitioned QCA (PQCA) are intrinsically universal. Since they are the most canonical description of QCA, we will, without loss of generality, assume that all QCA are PQCA throughout this work.
\begin{Def}[Partitioned QCA]\label{def:pqca}
A partitioned  $n$-dimensional quantum cellular automaton (PQCA) is defined by a scattering unitary operator $U$ such that $U:\mathcal{H}_{\Sigma}^{\otimes 2^n}\longrightarrow\mathcal{H}_{\Sigma}^{\otimes 2^n}$, and $U\ket{qq\ldots qq}=\ket{qq\ldots qq}$, \ie that takes
a hypercube of $2^n$ cells into a hypercube of $2^n$ cells and preserve quiescence. Consider $G=(\bigotimes_{2\mathbb{Z}^n} U)$, the operator over $\mathcal{H}$. The induced global evolution is $G$ at odd time steps, and $\sigma G$ at even time steps, where $\sigma$ is a translation by one in all directions, see Fig. \ref{fig:structure}\VS{ \emph{(left.)}}.
\end{Def}
\VL{
\begin{figure}[h]
\centering
\includegraphics[scale=.9, clip=true, trim=0cm 0cm 0cm 0cm]{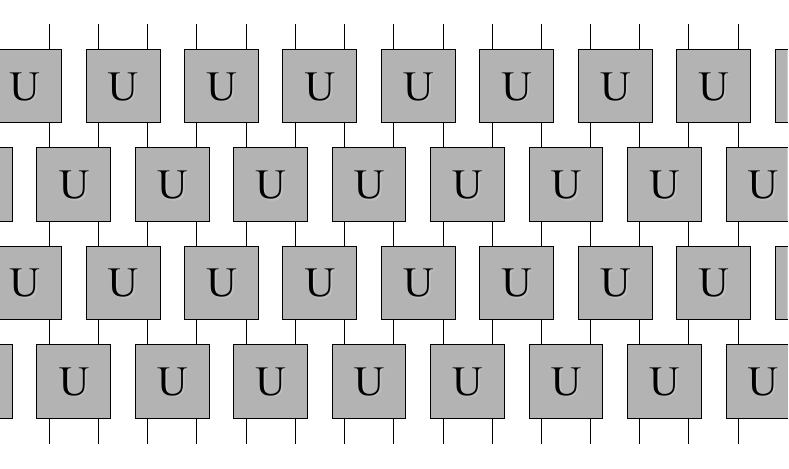}
\caption{\label{fig:structure} One-dimensional PQCA with scattering unitary $U$. Each line represents a quantum system, in this case a whole cell. Each square represents a scattering unitary $U$ which is applied to two cells. Time flows upwards.}
\end{figure}
}

\subsection{Intrinsic Simulation of $n$-Dimensional PQCA} \label{subsecsim}
\VL{Intrinsic simulation of one CA by another was discussed informally in section \ref{subsec:CA}. A
pedagogical discussion in the classical case is given in ref. \cite{OllingerJAC}.
Quantised intrinsic simulation was formalised in the one-dimensional case in ref. \cite{ArrighiUQCA}.}
The definition \VS{of intrinsic simulation} was extended to the quantum case in refs. \cite{ArrighiNUQCA,ArrighiPQCA}, with further discussion.
Here we simply recall these definitions.
\begin{Def}[Isometric coding]\label{isomcode}
Consider $\Sigma_G$ and $\Sigma_H$, two alphabets with distinguished quiescent states $q_G$ and $q_H$, and such that $|\Sigma_H|\leq|\Sigma_G|$. Consider $\mathcal{H}_{\Sigma_G}$ and $\mathcal{H}_{\Sigma_H}$ the Hilbert spaces having these alphabets as their basis, and $\mathcal{H}_{\mathcal{C}_f^{G}}$, $\mathcal{H}_{\mathcal{C}_f^{H}}$ the Hilbert spaces of finite configurations over these alphabets.\\
Let $E$ be an isometric linear map from $\mathcal{H}_{\Sigma_H}$ to $\mathcal{H}_{\Sigma_G}$ which preserves quiescence, \ie such that $E\ket{q_H}=\ket{q_G}$. It trivially extends into an isometric linear map $Enc=(\bigotimes_{\mathbb{Z}^n} E)$ from $\mathcal{H}_{\mathcal{C}_f^{H}}$ into $\mathcal{H}_{\mathcal{C}_f^{G}}$, which we call an isometric encoding.\\
Let $D$ be an isometric linear map from $\mathcal{H}_{\Sigma_G}$ to $\mathcal{H}_{\Sigma_H}\otimes\mathcal{H}_{\Sigma_G}$ which also preserves quiescence, in the sense that $D\ket{q_G}=\ket{q_H}\otimes\ket{q_G}$. It trivially extends into an isometric linear map $Dec=(\bigotimes_{\mathbb{Z}^n} D)$ from $\mathcal{H}_{\mathcal{C}_f^{G}}$ into $\mathcal{H}_{\mathcal{C}_f^{H}}\otimes\mathcal{H}_{\mathcal{C}_f^{G}}$, which we call an isometric decoding.\\
The isometries $E$ and $D$ define an isometric coding if the following condition is satisfied:\\
$\forall \ket{\psi}\in \mathcal{H}_{\mathcal{C}_f^{H}},\,\exists \ket{\phi}\in \mathcal{H}_{\mathcal{C}_f^{G}}\quad/\quad\ket{\psi}\otimes\ket{\phi}=Dec\pa{Enc \ket{\psi}}.$
\end{Def}
(Here $Dec$ is understood to morally be an inverse function of $Enc$, but some garbage $\ket{\phi}$ may be omitted.)
\begin{Def}[Direct simulation]\label{directsim}
Consider $\Sigma_G$ and $\Sigma_H$, two alphabets with distinguished quiescent states $q_G$ and $q_H$, and two QCA $G$ and $H$ over these alphabets. We say that $G$ directly simulates $H$, if and only if there exists an isometric coding such that\\
$\forall i\in\mathbb{N},\,\forall \ket{\psi}\in \mathcal{H}_{\mathcal{C}_f^{H}},\,\exists \ket{\phi}\in \mathcal{H}_{\mathcal{C}_f^{G}}\quad/\quad (G^i\ket{\psi})\otimes\ket{\phi}=Dec \pa{{H}^i\pa{Enc \ket{\psi}}}.$
\end{Def}
\begin{Def}[Grouping]\label{def:packmap}
Let $G$ be an $n$-dimensional QCA over alphabet $\Sigma$. Let $s$ and $t$ be two integers, $q'$ a word in $\Sigma'=\Sigma^{s^n}$. Consider the iterate global evolution $G^t$ up to a grouping of each hypercube of $s^n$ adjacent cells into one supercell. If this operator can be considered to be a QCA $G'$ over $\Sigma'$ with quiescent symbol $q'$, then we say that $G'$ is an $(s,t,q')$-grouping of $G$.
\end{Def}
\begin{Def}[Intrinsic simulation]\label{def:intsim}
Consider $\Sigma_G$ and $\Sigma_H$, two alphabets with distinguished quiescent states $q_G$ and $q_H$, and two QCA $G$ and $H$ over these alphabets. We say that $G$ intrinsically simulates $H$ if and only if there exists $G'$ some grouping of $G$ and $H'$ some grouping of $H$ such that $G'$ directly simulates $H'$.
\end{Def}
In other words, $G$ intrinsically simulates $H$ if and only if there exists some isometry $E$ which translates supercells of $H$ into supercells of $G$, such that if $G$ is iterated and then translated back, the whole process is equivalent an iteration of $H$,  as in Fig. \VL{\ref{UsimV}}\VS{\ref{fig:structure} \emph{(right)}}.
\VL{
\begin{figure}[h]
\centering
\includegraphics[scale=0.75, clip=true, trim=0cm 0cm 0cm 0cm]{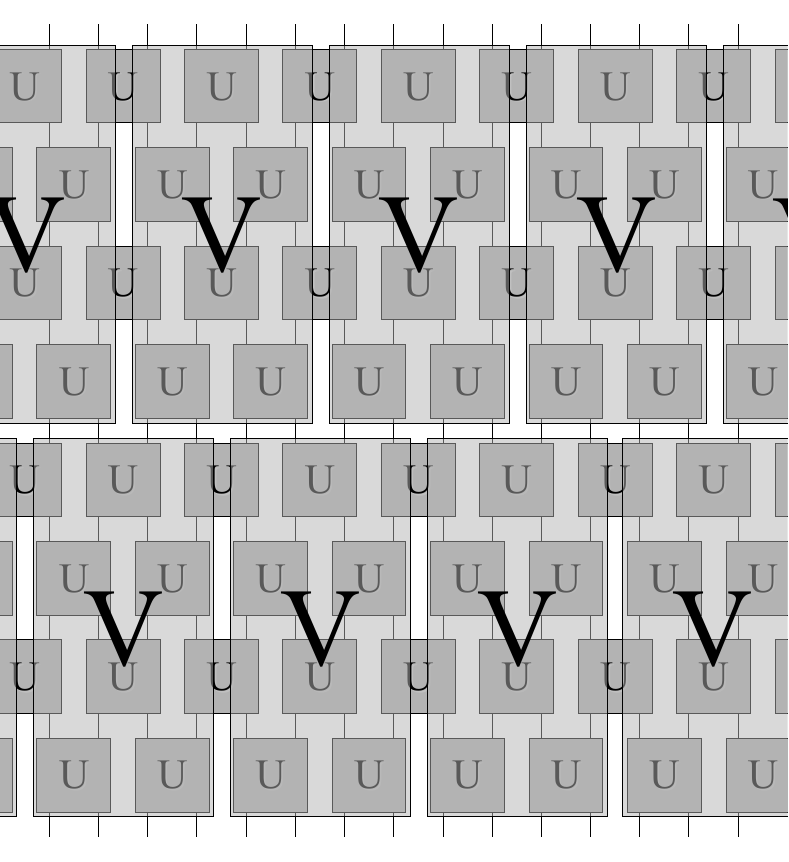}
\caption{Intrinsic simulation of one PQCA by another.\label{UsimV}
The PQCA defined by $U$ simulates the PQCA defined by $V$. In this case, two cells of the $U$-defined PQCA are required to encode one cell of the $V$-defined PQCA, and the $U$-defined PQCA executes four time steps to simulate one time step of the $V$-defined PQCA. }
\end{figure}
}
\VS{
\begin{figure}[h]
\centering
\VS{\vspace{-10mm}}\includegraphics[scale=0.8, clip=true, trim=0cm 0cm 0cm 0cm]{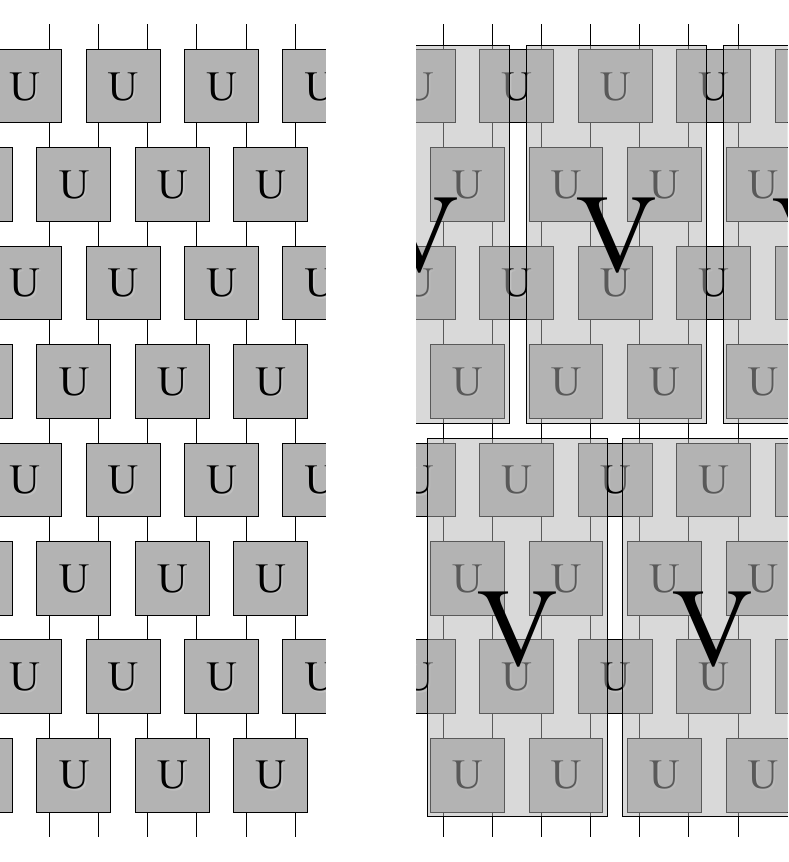}
\caption{\label{fig:structure} \label{UsimV} \emph{Left:} Partitioned one-dimensional PQCA with scattering unitary $U$. Each line represents a quantum system, in this case a whole cell. Each square represents a scattering unitary $U$ which is applied to two cells. Time flows upwards. \emph{Right:} The PQCA defined by $U$ simulates the PQCA defined by $V$. In this case two cells of the $U$-defined PQCA encode one cell of the $V$-defined PQCA, and the $U$-defined PQCA is run for four time steps to simulate one time step of the $V$-defined PQCA. \VL{More generally the challenge is to define an initial configuration of the $U$-defined PQCA so that it behaves just as the $V$-defined PQCA with respect to the encoded initial configuration, after some fixed number of time steps. Clearly such an encoding must hold the configuration of the $V$-defined PQCA as well as a way of describing the scattering unitary $V$.}}
\end{figure}
}
} 

\section{An Intrinsically Universal QCA}\label{sec:nuqca}

\VD{In section \ref{subsecdef} the formal definition of $n$-dimensional PQCA was discussed (Fig. \ref{fig:structure}), and the formal definition of intrinsic simulation was recalled (Fig. \ref{UsimV}). }
The aim \VD{now }is to find a particular $U$-defined PQCA which is capable of intrinsically simulating any $V$-defined PQCA, for any $V$.
In order to describe such a $U$-defined PQCA in detail, two things must be given: 
the dimensionality of the cells (including the meaning attached to each of the states they may take), and the way the scattering unitary $U$ acts upon these cells.
\VL{First we discuss the general scheme used to solve this problem, and then describe the PQCA implementing it.} \VND{The necessary definitions for $n$-dimensional QCA are given in refs. \cite{ArrighiNUQCA,ArrighiPQCA}.}
\VS{\medskip}

\subsection{Circuit Universality versus Intrinsic Universality in Higher Dimensions}
\VS{\indent} As already discussed, intrinsic universality refers to the ability for one CA to simulate any other CA\VS{,} \VL{in a way which preserves the spatial structure of the simulated CA.
Conversely, computation universality refers to the ability of a CA to simulate any TM, and hence run any algorithm.}\VS{whereas computation universality is about simulating a TM.} Additionally, circuit universality is the ability of one CA to simulate any circuit. \VL{These are \textsc{Nand} gate circuits for classical circuits and CA, and \textsc{Toffoli} gate circuits for reversible circuits and CA.} Informally, in a quantum setting, circuit universality is the ability of a PQCA to simulate \VL{any unitary evolution expressed as a}\VS{any finitary} combination of a universal set of quantum gates, such as the standard gate set: \textsc{Cnot, \Phase} (also known as the $\frac{\pi}{8}$ gate), and the \textsc{Hadamard} gate.
\VL{
The relationships between these three concepts of CA universality have been noted previously \cite{DurandRoka}.
A computation universal CA is also a circuit universal CA, because circuits are finitary computations.
Moreover, an intrinsic universal CA is also a computation universal CA, because it can simulate any CA,
including computation universal CA. Hence intrinsic universality implies computation universality, which implies circuit universality.

In one-dimension this is not an equivalence. Intuitively, computation universality requires more than circuit universality, namely the ability to loop the computation, which is not trivial for CA. Similarly, intrinsic universality requires more than computation universality, such as the ability to simulate multiple communicating TM. In the classical setting there are formal results that distinguish these ideas \cite{OllingerJAC}.}

In $n$-dimensions, it is often assumed in the classical CA literature that circuit universality implies intrinsic universality, and \VL{hence these are all equivalent}\VS{that both are equivalent to computation universality} \cite{OllingerJAC}\VS{, without provision of an explicit construction}. Strictly speaking this is not true. Consider a two-dimensional CA which runs one-dimensional CA in parallel. If the one-dimensional CA is circuit/computation universal, but not computation/intrinsically universal, then this is also true for the two-dimensional CA. Similarly, in the PQCA setting, the two-dimensional constructions in \cite{PerezCheung} and \cite{Raussendorf} are circuit universal but not intrinsically universal.

However, this remains a useful intuition: Indeed, CA admit a block representation, where these blocks are permutations for reversible CA, while for PQCA the blocks are unitary matrices. Thus the evolution of any (reversible/quantum) CA can be expressed as an infinite (reversible/quantum) circuit of (reversible/ quantum) gates repeating across space. If a CA is circuit universal, and if it is possible to wire together different circuit components in different regions of space, then the CA can simulate the block representation of any CA, and hence can simulate any CA in a way which preserves its spatial structure. It is intrinsically universal.\VS{\medskip}
\VL{
This is the route followed next in constructing the intrinsically universal $n$-dimensional PQCA. First the construction of  the `wires',
which can carry information across different regions of space, is considered. Here these are signals which can be redirected or delayed using barriers, with each signal holding a qubit of information. Secondly, the `circuit-pieces' are constructed, by implementing quantum gates which can be combined. One and two qubit gates are implemented as obstacles to, and interactions of, these signals.}

\subsection{Flattening a PQCA into Space}\label{subsec:flat}
\VL{In the classical CA literature it is considered enough to show that the CA implements some wires carrying signals, and some universal gates
acting upon them, to prove that an $n$-dimensional CA is in fact intrinsically universal. }
\VS{\indent}Any CA can be encoded into a `wire and gates' arrangement following the above argument, but this has never been made explicit in the literature.
This section makes more precise how to flatten any PQCA in space, so that it is simulated by a PQCA which implements quantum
wires and universal quantum gates. Flattening a PQCA means that the infinitely repeating, two-layered circuit is arranged in space so that at the beginning all the signals carrying qubits find themselves in circuit-pieces which implement a scattering unitary of the first layer, and then all synchronously exit and travel to circuit-pieces implementing the scattering unitary of the second layer, etc.
An algorithm for performing this flattening can be provided, however the process will not be described in detail,
for clarity and following the classical literature, which largely ignores this process.

\begin{figure}
\centering
\VS{\includegraphics[scale=.85, clip=true, trim=0cm 0cm 0cm 0cm]{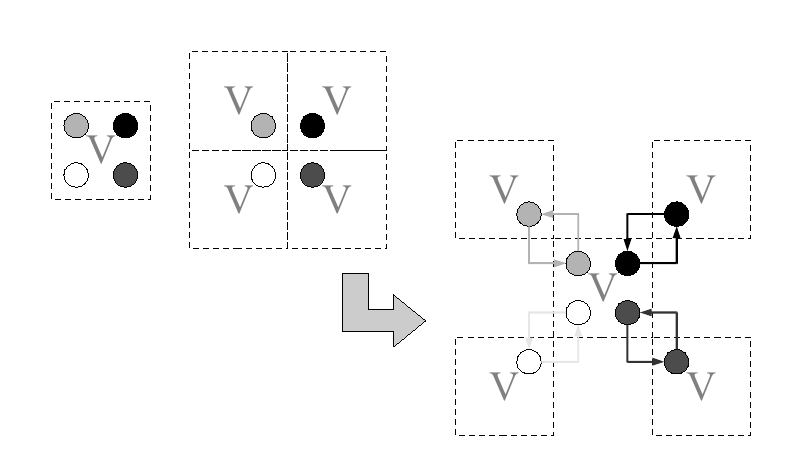}}
\VL{\includegraphics[scale=.9, clip=true, trim=0cm 0cm 0cm 0cm]{img/flattening1and2.pdf}}
\caption{Flattening a PQCA into a simulating PQCA. \emph{Left}: Consider four cells (white, light grey, dark grey, black) of a PQCA having scattering unitary $V$.
The first layer PQCA applies $V$ to these four cells, then the second layer applies $V$ at the four corners. \emph{Right}: We need to flatten this so that the two-layers become non-overlapping. The first layer corresponds to the centre square, and the second layer to the four corner squares. At the beginning the signals (white, light grey, dark grey, black) coding for the simulated cells are in the centre square. \VL{They undergo $V$, and are directed towards the bottom left, top left, bottom right, and top right squares respectively, where they undergo $V$ but paired up with some other signals, etc.}
\label{fig:flattening12}}
\end{figure}

\begin{figure}
\centering
\VS{\includegraphics[scale=.8, clip=true, trim=0cm 0cm 0cm 0cm]{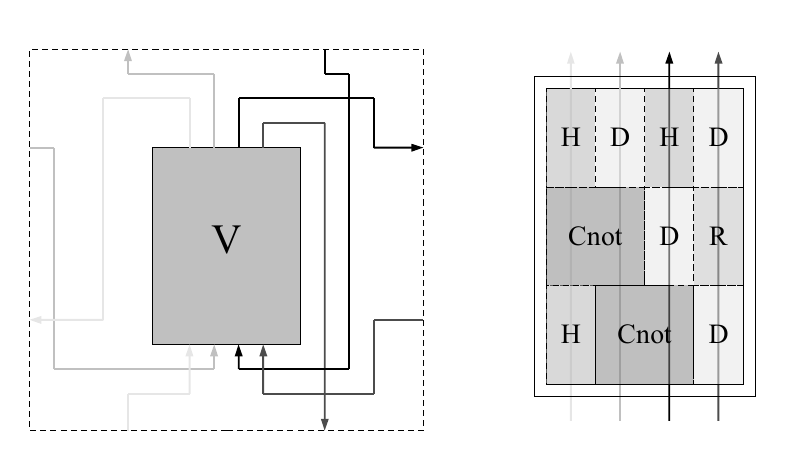}}
\VL{\includegraphics[scale=.9, clip=true, trim=0cm 0cm 0cm 0cm]{img/flattening3and4.pdf}}
\caption{Flattening a PQCA into a simulating PQCA (cont'd). \emph{Left}: Within the central square the incoming signals are bunched together so as to undergo a circuit which implements $V$, and are then dispatched towards the four corners. This diagram does not make explicit a number of signal delays, which may be needed to ensure that they arrive synchronously at the beginning of the circuit implementing $V$. \emph{Right}: Within the central rectangle, the circuit which implements $V$ is itself a combination of smaller circuits for implementing a universal set of quantum gates such as \textsc{Cnot}, \textsc{Hadamard} and the \textsc{\Phase}, together with delays. \VL{These are implemented as explained in sections \ref{subsec:onequbit} and \ref{subsec:gates}.}\label{fig:flattening34}}
\end{figure}

The flattening process can be expressed in three steps:
Firstly, the $V$-defined PQCA is expanded in space by coding each cell into a hypercube of $2^n$ cells. This allows enough space for the scattering unitary $V$ to be applied on non-overlapping hypercubes of cells,  illustrated in the two-dimensional case in Fig.~\ref{fig:flattening12}.
\noindent Secondly, the hypercubes where $V$ is applied must be connected with wires, as shown in Fig. \ref{fig:flattening12} $(right)$. Within these hypercubes wiring is required so that incoming signals are bunched together to undergo a circuit implementation of $V$, and are then dispatched appropriately, as shown in Fig. \ref{fig:flattening34} $(left)$. This requires both time and space expansions, with factors that depend non-trivially (but uninterestingly) upon the size of the circuit implementation of $V$ and the way the wiring and gates work in the simulating PQCA.
\noindent Next, an encoding of the circuit description of the scattering unitary $V$ is implemented in the simulating PQCA upon these incoming bunched wires, as shown in Fig. \ref{fig:flattening34} $(right)$. This completes the description of the overall scheme according to which a PQCA that is capable of implementing wires and gates is also capable of intrinsically simulating any PQCA, and hence any QCA. A particular PQCA that supports these wires and gates can now be constructed.\VS{\newpage}

\subsection{Barriers and Signals Carrying Qubits}\label{subsec:onequbit}
\VS{\indent} Classical CA studies often refer to `signals' without an explicit definition. In this context,
a signal refers to the state of a cell which may move to a neighbouring cell consistently, from one step to another, by the evolution of the CA.
Therefore a signal would appear as a line in the space-time diagram of the CA. These lines need to be implemented as signal redirections.
A $2$D solution is presented here, but this scheme can easily be extended to higher dimensions. Each cell has four possible basis states:
empty ($\epsilon$), holding a qubit signal ($0$ or $1$), or a barrier ($\blacksquare$). The scattering unitary $U$ of the universal PQCA acts on $2\times 2$ cell neighbourhoods.

Signals encode qubits which can travel diagonally across the 2D space (NE, SE, SW, or NW).
Barriers do not move, while signals move in the obvious way if unobstructed, as there is only one choice for any signal in any square of four cells.
Hence the basic movements of signals are given by the following four rules:
$$\cells{}{}{$s$}{} \mapsto \cells{}{$s$}{}{}, \qquad \cells{$s$}{}{}{} \mapsto \cells{}{}{}{$s$},$$
$$\cells{}{$s$}{}{} \mapsto \cells{}{}{$s$}{}, \qquad \cells{}{}{}{$s$} \mapsto \cells{$s$}{}{}{}.$$
where $s\in \{0,1\}$ denotes a signal, and blank cells are empty.

The way to interpret the four above rules in terms of the scattering unitary $U$ is just case-by-case definition, \ie they show that
$U\cells{}{}{$s$}{}=\cells{}{$s$}{}{}$.
Moreover, each rule can be obtained as a rotation of another, hence by stating that the $U$-defined PQCA is isotropic the first rule above suffices. This convention will be used throughout.

\begin{figure}
\centering
\VS{\includegraphics[scale=.55, clip=true]{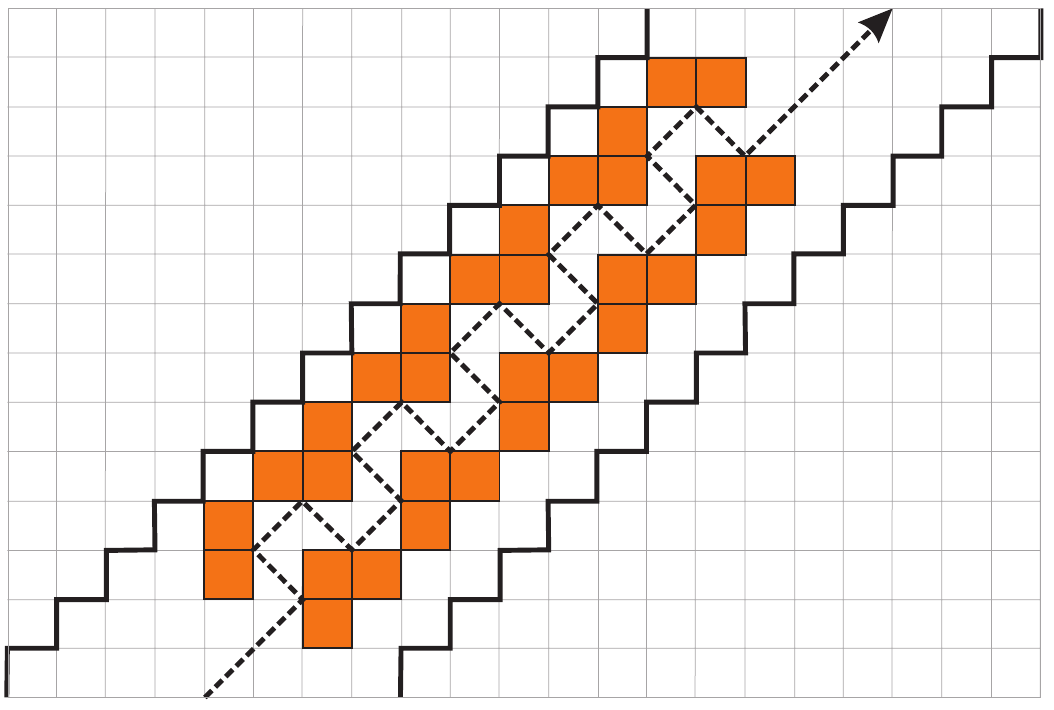}}
\VL{\includegraphics[scale=.60, clip=true]{img/delayCirc.pdf}}
\caption{The `identity circuit' tile, an $8\times 14$ tile taking 24 time-steps, made by repeatedly bouncing the signal
from walls to slow its movement through the tile. The dotted line gives the signal trajectory, with the arrow showing the
exit point and direction of signal propagation. The bold lines show the tile boundary.}
\label{fig:delays}
\end{figure}

\begin{figure}
\centering
\VS{\includegraphics[scale=.55, clip=true]{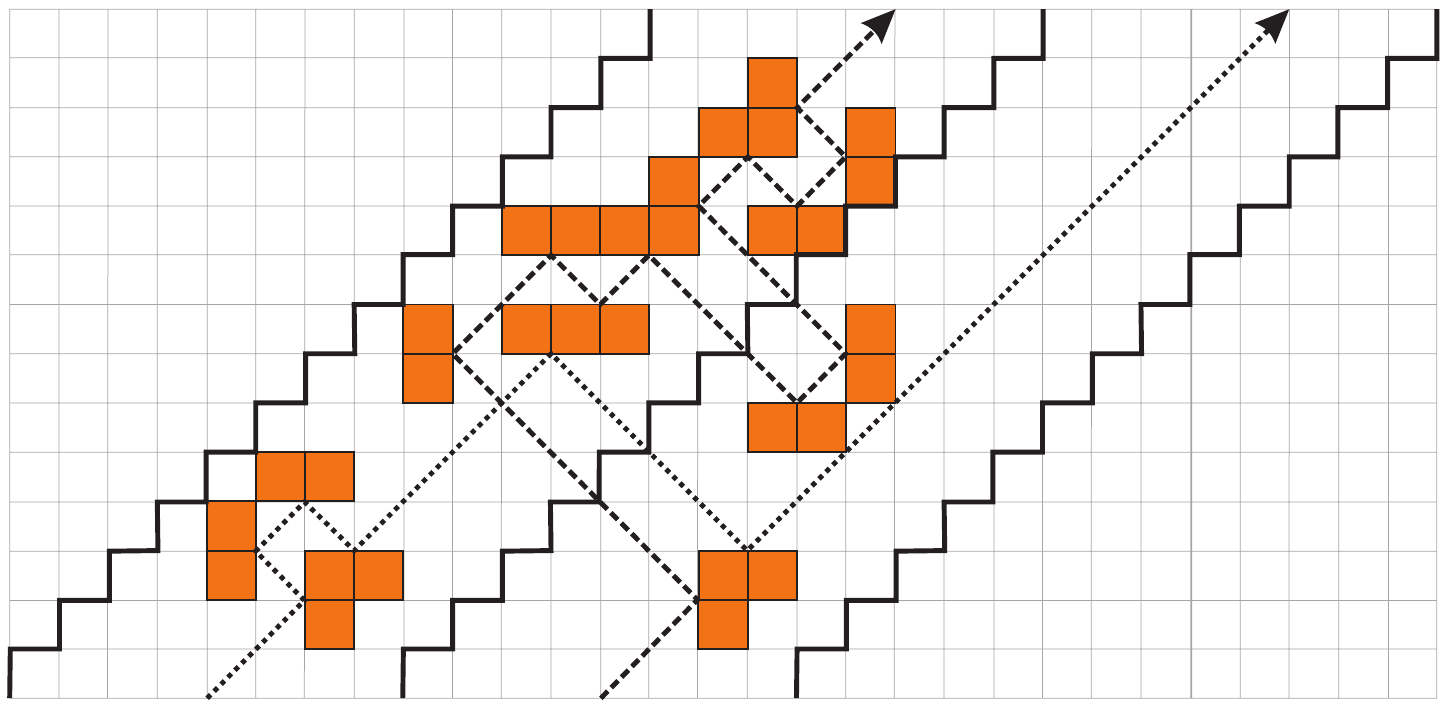}}
\VL{\includegraphics[scale=.60, clip=true]{img/swapCirc.pdf}}
\caption{The `swap circuit' tile, a $16\times 14$ tile, where both input signals are permuted and exit synchronously after 24 time-steps.
As the first signal (\emph{bottom left}) is initially delayed, there is no interaction.}
\label{fig:swap}
\end{figure}

The ability to redirect signals is achieved by `bouncing' them off walls constructed from
two barriers arranged either horizontally or vertically:
$$
\cells{\barrier}{$s$}{\barrier}{} \mapsto \cells{\barrier}{}{\barrier}{$s$}.
$$
where $s$ again denotes the signal and the shaded cells denote the barriers which causes the signal
to change direction.
If there is only one barrier present in the four cell square being operated on then the signal simply propagates as normal and is not deflected:
$$\cells{\barrier}{}{$s$}{} \mapsto \cells{\barrier}{$s$}{}{}.$$
Using only these basic rules of signal propagation and signal reflection from barrier walls, signal delay (Fig.  \ref{fig:delays}) and signal swapping (Fig. \ref{fig:swap})
tiles can be constructed. All of the rules presented so far are permutations of some of the base elements of the vector space generated by
$$\Set{\cells{$w$}{$x$}{$y$}{$z$}}_{w,x,y,z \in \{\epsilon,0,1,\blacksquare\}}$$
therefore $U$ is indeed unitary on the subspace upon which its action has so far been described.

\subsection{Gates}\label{subsec:gates}
\begin{figure}
\centering
\VS{\vspace{-0.5mm}\includegraphics[scale=.55, clip=true]{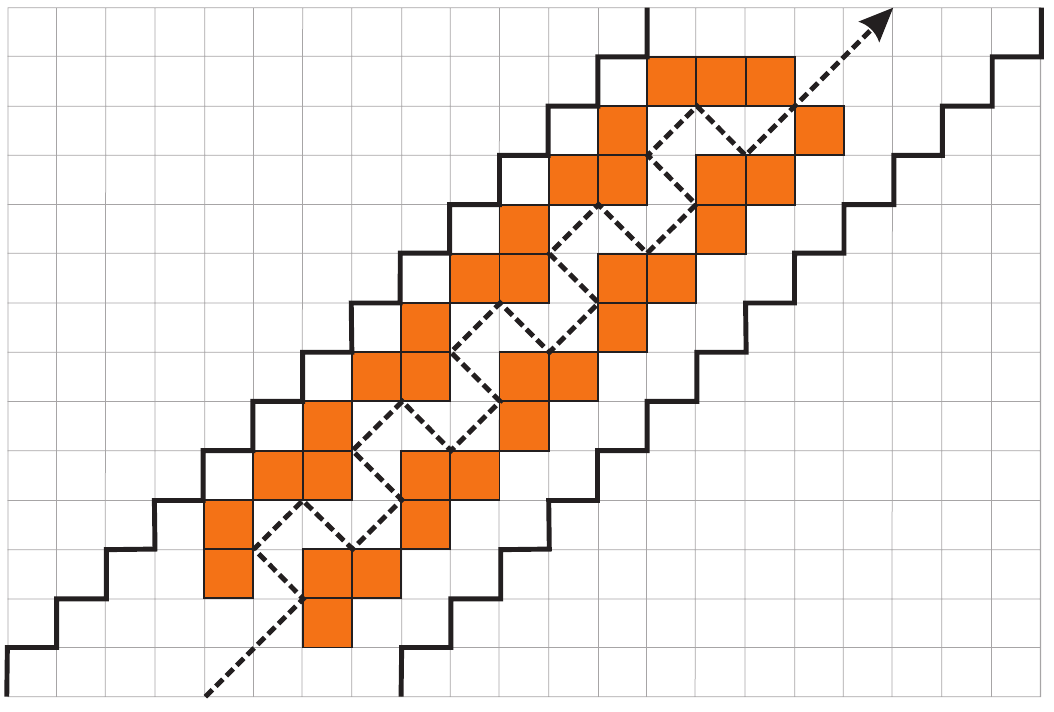}}
\VL{\includegraphics[scale=.60, clip=true]{img/hadCirc.pdf}}
\caption{The `Hadamard gate' tile applies the Hadamard operation to the input signal. It is a modification
of the identity circuit tile, with a diagonal (semitransparent) barrier added at the end which performs the Hadamard operation.}
\label{fig:hadamard}
\end{figure}

\begin{figure}
\centering
\VS{\vspace{-0.5mm}\includegraphics[scale=.55, clip=true]{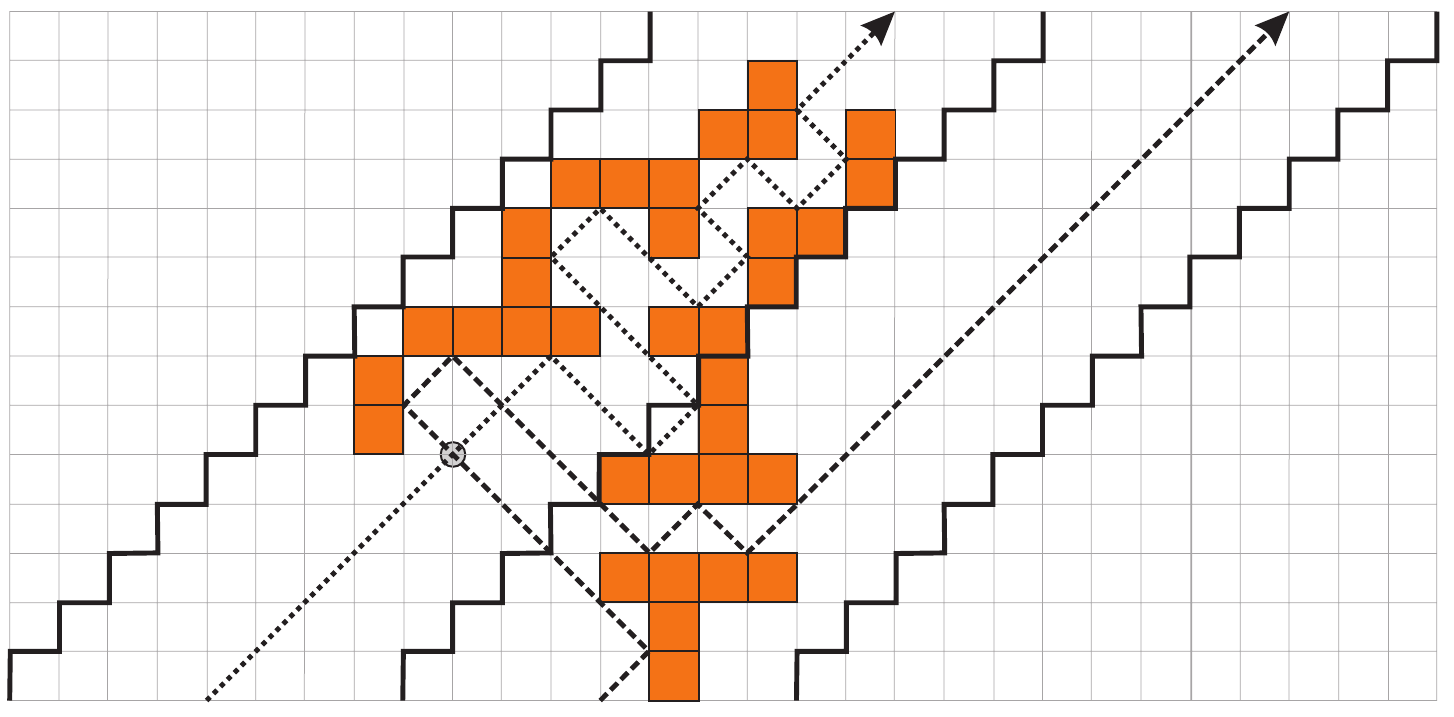}}
\VL{\includegraphics[scale=.60, clip=true]{img/cPhaseCirc.pdf}}
\caption{The `\cPhase{} gate' tile\VS{, with a signal interaction at the highlighted cell.}\VL{ applies the controlled-\Phase{} operation to the two input qubits, by causing the signals to interact at the highlighted point (grey circle). The qubits are then synchronised so that they exit at the same time along their original paths. No swapping takes place.}}
\label{fig:cphase}
\end{figure}

\begin{figure}
\centering
\VS{\vspace{-0.5mm}\includegraphics[scale=.55, clip=true]{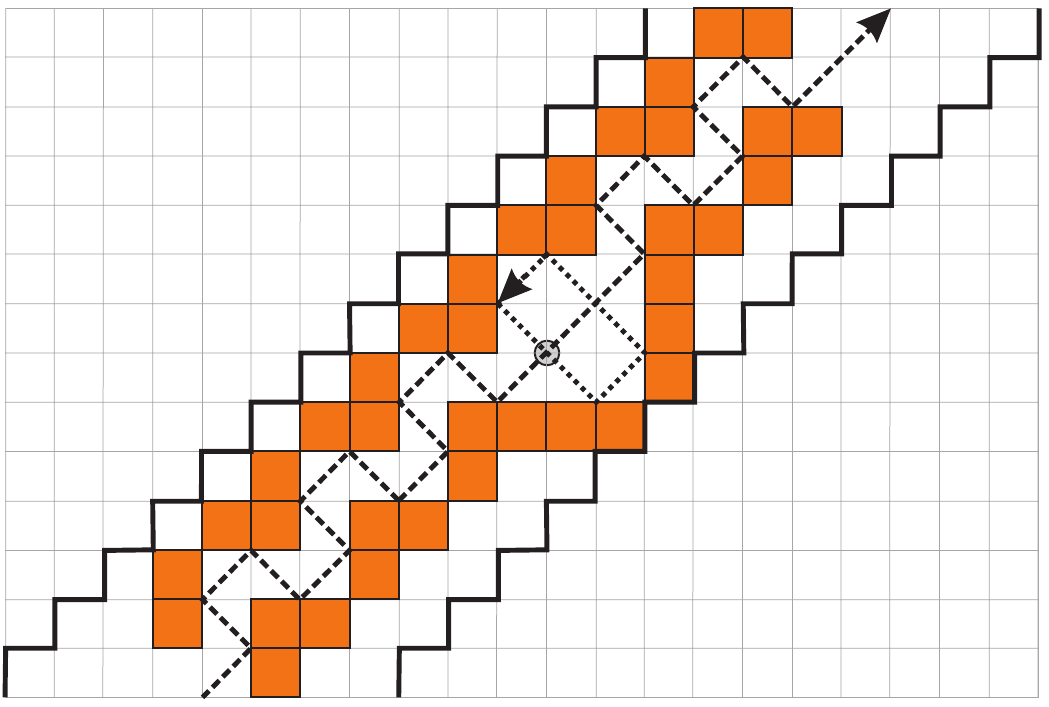}}
\VL{\includegraphics[scale=.60, clip=true]{img/phaseCirc.pdf}}
\caption{The `\Phase{} gate' tile. This tile makes use
of a signal, set to $\ket{1}$,  which loops inside the grid every six time-steps, ensuring that it will interact with the signal that enters the
tile, and causing it to act as the control qubit to a \cPhase{} operation. It therefore acts as a phase rotation on the input qubit, which passes directly through. \VL{After 24 time-steps the auxiliary control signal has returned to its origin, unchanged, hence the tile can be reused.}}
\label{fig:phase}
\end{figure}

\VS{\indent} To allow a universal set of gates to be implemented by the PQCA,  certain combinations
of signals and barriers can be assigned special importance.
The Hadamard operation on a single qubit-carrying signal can be implemented by interpreting a
signal passing through a diagonally oriented wall, analogous to a semitransparent barrier in physics. This has the action
defined by the following rule:
$$\cells{\barrier}{}{0}{\barrier} \mapsto\frac{1}{\sqrt{2}}\cells{\barrier}{0}{}{\barrier} + \frac{1}{\sqrt{2}}\cells{\barrier}{1}{}{\barrier}$$
$$\cells{\barrier}{}{1}{\barrier} \mapsto \frac{1}{\sqrt{2}}\cells{\barrier}{0}{}{\barrier} - \frac{1}{\sqrt{2}}\cells{\barrier}{1}{}{\barrier} $$
This implements the Hadamard operation, creating a superposition of configurations with appropriate phases. Using this construction
a Hadamard tile can be constructed (Fig. \ref{fig:hadamard}) by simply adding a semitransparent barrier to the end of the previously
defined delay (identity) tile (Fig. \ref{fig:delays}).
A way of encoding two qubit gates in this system is to consider that two signals which
cross paths interact with one another. The controlled-\Phase{ }
operation can be implemented by considering signals that cross each other as interacting only if they are both $1$, in which case a global phase
of $e^{\frac{i\pi}{4}}$ is applied. Otherwise the signals continue as normal. This behaviour is defined by the following rule:
$$\cells{1}{}{1}{} \mapsto e^{\frac{i\pi}{4}}\cells{}{1}{}{1}, \qquad \cells{$x$}{}{$y$}{} \mapsto \cells{}{$y$}{}{$x$} otherwise$$
where $x,y \in \{0,1\}$. This signal interaction which induces a global phase change allows the definition of both a two signal
controlled-\Phase{} tile (Fig. \ref{fig:cphase}) and a single signal \Phase{} operation tile (Fig. \ref{fig:phase}).
These rules are simply a permutation and phase change of base elements of the form:
$$\Set{\cells{$x$}{}{$y$}{}}_{x,y \in \{0,1\}}$$
(and their rotations), therefore $U$ is a unitary operation on the subspace upon which its action has so far been described.
Wherever $U$ has not yet been defined, it is the identity. Hence $U$ is unitary.\VS{\medskip}

\subsection{Circuits: Combining Gates}\label{subsec:circuits}
\VS{\indent} A signal is given an $8 \times 14$ tile ($16 \times 14$ for two signal operations)
in which the action is encoded. The signals enter each tile at the fifth cell from the left, and propagate diagonally NE.
Each time step finds the tile shifted one cell to the right to match this diagonal movement, giving a diagonal tile.
The signal exits the tile $14$ cells North and East of where it entered. This allows
these tiles to be composed in parallel and sequentially with the only other requirement being that the signal exits at the appropriate point, \ie
the fifth cell along the tile, after $24$ time-steps. This ensures that all signals are synchronised as in Fig. \ref{fig:flattening34} (\emph{right}),
allowing larger circuits to be built from these elementary tiles by simply plugging them together. Non-contiguous gates can also be wired together
using appropriate wall constructions to redirect and delay signals so that they are correctly synchronised.

The implemented set of quantum gates, the identity, Hadamard, swap, \Phase{} and \cPhase{}, gives a universal set. Indeed the standard set
of \textsc{cNot}, \textsc{H}, \textsc{\Phase} can be recovered as follows:
$$\textsc{cNot}\ket{\psi}=(\mathbb{I}\otimes H)(\textsc{cR(${\pi}\slash{4}$)})^4(\mathbb{I}\otimes H)\ket{\psi}$$
where $\textsc{cR($\frac{\pi}{4}$})^4$ denotes four applications of the \cPhase{} gate, giving the controlled-\textsc{Phase} operation.

\VS{\vspace{-0.5mm}}\section{Conclusion}\label{sec:discussion}

This paper presents a simple PQCA which is capable of simulating all other PQCA, preserving the topology
of the simulated PQCA. This means that the initial configuration and the forward evolution of any PQCA can be encoded
within the initial configuration of this PQCA, with each simulated cell encoded as a group of adjacent cells in the PQCA,
\ie intrinsic simulation.
The construction in section \ref{sec:nuqca} is given in two-dimensions, which can be seen to generalise to $n>1$-dimensions.
The main, formal result of this work can therefore be stated as:
\begin{Cl}
There exists an $n$-dimensional $U$-defined PQCA, $G$, which is an intrinsically universal PQCA. Let $H$ be a $n$-dimensional $V$-defined PQCA such that $V$ can be expressed as a quantum circuit $C$ made of gates from the set $\textsc{Hadamard}$, $\textsc{Cnot}$, and $\textsc{\Phase}$. Then $G$ is able to intrinsically simulate $H$.
\end{Cl}

Any finite-dimensional unitary $V$ can always be approximated by a circuit $C(V)$ with an arbitrary small error $\varepsilon=\max_{\ket{\psi}}||V\ket{\psi}-C\ket{\psi}||$. Assuming instead that $G$ simulates the $C(V)$-defined PQCA, for a region of $s$ cells over a period $t$, the error with respect to the $V$-defined PQCA will be bounded by $st\varepsilon$. This is due to the general statement that errors in quantum circuits increase, at most, proportionally with time and space \cite{NielsenChuang}.
Combined with the fact that PQCA are universal \cite{ArrighiNUQCA,ArrighiPQCA}, this means that $G$ is intrinsically universal, up to this unavoidable approximation.\VS{\medskip}

\subsection{Discussion and Future Work}
\indent QC research has so far focused on applications for more secure and efficient computing, with theoretical physics supporting this work in theoretical computer science. The results of this interdisciplinary exchange led to the assumptions underlying computer science being revisited,
with information theory and complexity theory, for example, being reconsidered and redeveloped.
However, information theory also plays a crucial role in the foundations of theoretical physics\VS{ (\eg deepening our understanding of entanglement \cite{DurVidalCirac} and decoherence \cite{PazZurek})}.
These developments are also of interest in theoretical physics studies where physical aspects such as particles and matter are considered; computer science studies tend to consider only abstract mathematical quantities.
Universality, among the many computer science concepts, is a simplifying methodology in this respect. For example, if the problem being studied crucially involves some idea of interaction, universality makes it possible cast it in terms of information exchanges \emph{together} with some universal information processing.
This paper presents an attempt to export universality as a tool for application in theoretical physics, a small step towards the goal of finding and understanding a  \emph{universal physical phenomenon}, within some simplified mechanics.
Similar to the importance of the idea of the spatial arrangement of interactions in physics, intrinsic universality has broader applicability than computation universality and must be preferred. In short, if only one physical phenomenon is considered, it should be an intrinsically universal physical phenomenon, as it could be used to simulate all others.

The PQCA cell dimension of the simple intrinsically universal construction given here is four (empty, a qubit ($\ket{0}$ or $\ket{1}$), or a barrier). In comparison,
the simplest classical Partitioned CA has cell dimension two \cite {MargolusQCA}.
Hence, although the intrinsically universal PQCA presented here is the simplest found, it is not minimal. In fact, one can also manage \cite{Arrigh2UQCA} an intrinsically universal PQCA with a cell dimension of three, in two different ways. One way is to encode the spin degree of freedom  (0 and 1) into a spatial degree of freedom, so that now the semitransparent barrier either splits or combines signals.
The second way is to code barriers as pairs of signals as in the Billiard Ball CA model \cite{MargolusQCA}.
These constructions may be minimal, but are not as elegant as the one presented here.
In future work we will show that there is a simple, greater than two-dimensional PQCA which is minimal, as it has a cell dimension of two.

\section*{Acknowledgements}
The authors would like to thank  J\'er\^ome Durand-Lose, Jarkko Kari, Jacques Mazoyer, Kenichi Morita, Nicolas Ollinger, Guillaume Theyssier and Philippe Jorrand.

\bibliography{biblio}

\begin{thebibliography}{10}

\bibitem{AlbertCulik}
Albert, J., Culik, K.:
\newblock {A simple universal cellular automaton and its one-way and totalistic
  version}.
\newblock Complex Systems \textbf{1} (1987)  1--16

\bibitem{ArrighiMFCS}
Arrighi, P.:
\newblock {Algebraic characterizations of unitary linear quantum cellular
  automata}.
\newblock In: Proceedings of MFCS, Lecture Notes in Computer Science. Volume
  4162., Springer (2006)  122

\bibitem{ArrighiUQCA}
Arrighi, P., Fargetton, R.:
\newblock {Intrinsically universal one-dimensional quantum cellular automata}.
\newblock In: Proceedings of the {Development of Computational Models} workshop
  (DCM '07). (2007)

\bibitem{ArrighiFI}
Arrighi, P., Fargetton, R., Wang, Z.:
\newblock {Intrinsically universal one-dimensional quantum cellular automata in
  two flavours}.
\newblock Fundamenta Informaticae \textbf{21} (2009)  1001--1035

\bibitem{ArrighiNUQCA}
Arrighi, P., Grattage, J.:
\newblock {Intrinsically universal $n$-dimensional quantum cellular automata}.
\newblock Extended version of this paper. ArXiv preprint: arXiv:0907.3827
  (2009)

\bibitem{ArrighiPQCA}
Arrighi, P., Grattage, J.:
\newblock {Partitioned quantum cellular automata are intrinsically universal}.
\newblock Submitted (2009)

\bibitem{Arrigh2UQCA}
Arrighi, P., Grattage, J.:
\newblock {Two minimal $n$-dimensional intrinsically universal quantum cellular
  automata}.
\newblock Manuscript (2009)

\bibitem{ArrighiUCAUSAL}
Arrighi, P., Nesme, V., Werner, R.:
\newblock {Unitarity plus causality implies localizability}.
\newblock {Quantum Information Processing (QIP)} 2010, ArXiv preprint:
  arXiv:0711.3975 (2007)

\bibitem{ArrighiLATA}
Arrighi, P., Nesme, V., Werner, R.F.:
\newblock {Quantum cellular automata over finite, unbounded configurations}.
\newblock In: Proceedings of MFCS, Lecture Notes in Computer Science. Volume
  5196., Springer (2008)  64--75

\bibitem{Banks}
Banks, E.R.:
\newblock {Universality in cellular automata}.
\newblock In: Proceedings of the 11th Annual Symposium on Switching and
  Automata Theory (SWAT '70), Washington, DC, USA, IEEE Computer Society (1970)
   194--215

\bibitem{BrennenWilliams}
Brennen, G.K., Williams, J.E.:
\newblock {Entanglement dynamics in one-dimensional quantum cellular automata}.
\newblock Phys. Rev. A \textbf{68}(4) (Oct 2003)  042311

\bibitem{DurandRoka}
Durand, B., Roka, Z.:
\newblock {The Game of Life: universality revisited, Research Report 98-01}.
\newblock Technical report, Ecole Normale Suprieure de Lyon (1998)

\bibitem{Durand-LoseLATIN}
Durand-Lose, J.O.:
\newblock {Reversible cellular automaton able to simulate any other reversible
  one using partitioning automata}.
\newblock In: Latin '95, number 911, Lecture Notes in Computer Science,
  Springer (1995)  23024--4

\bibitem{Durand-LoseIntrinsic1D}
Durand-Lose, J.O.:
\newblock {Intrinsic universality of a 1-dimensional reversible cellular
  automaton}.
\newblock In: Proceedings of STACS '97, Lecture Notes in Computer Science,
  Springer (1997)  439

\bibitem{DurrWell}
Durr, C., Le~Thanh, H., Santha, M.:
\newblock {A decision procedure for well-formed linear quantum cellular
  automata}.
\newblock In: Proceedings of STACS '96, Lecture Notes in Computer Science,
  Springer (1996)  281--292

\bibitem{FeynmanQCA}
Feynman, R.P.:
\newblock {Quantum mechanical computers}.
\newblock Foundations of Physics (Historical Archive) \textbf{16}(6) (1986)
  507--531

\bibitem{LloydQG}
Lloyd, S.:
\newblock {A theory of quantum gravity based on quantum computation}.
\newblock ArXiv preprint: quant-ph/0501135 (2005)

\bibitem{MargolusPhysics}
Margolus, N.:
\newblock {Physics-like models of computation}.
\newblock Physica D: Nonlinear Phenomena \textbf{10}(1-2) (1984)

\bibitem{MargolusQCA}
Margolus, N.:
\newblock {Parallel quantum computation}.
\newblock In: Complexity, Entropy, and the Physics of Information: The
  Proceedings of the 1988 Workshop on Complexity, Entropy, and the Physics of
  Information, May-June 1989, Santa Fe, New Mexico, Perseus Books (1990)  273

\bibitem{MazoyerRapaport}
Mazoyer, J., Rapaport, I.:
\newblock {Inducing an order on cellular automata by a grouping operation}.
\newblock In: Proceedings of STACS '98, Lecture Notes in Computer Science,
  Springer (1998)  116--127

\bibitem{MoritaIntrinsicUniv1D}
Morita, K.:
\newblock {Reversible simulation of one-dimensional irreversible cellular
  automata}.
\newblock Theoretical Computer Science \textbf{148}(1) (1995)  157--163

\bibitem{MoritaCompUniv1D}
Morita, K., Harao, M.:
\newblock {Computation universality of one-dimensional reversible (injective)
  cellular automata}.
\newblock IEICE Trans. Inf. \& Syst., E \textbf{72} (1989)  758--762

\bibitem{MoritaCompUniv2D}
Morita, K., Ueno, S.:
\newblock {Computation-universal models of two-dimensional 16-state reversible
  cellular automata}.
\newblock IEICE Trans. Inf. \& Syst., E \textbf{75} (1992)  141--147

\bibitem{NagajWocjan}
Nagaj, D., Wocjan, P.:
\newblock {Hamiltonian Quantum Cellular Automata in 1D}.
\newblock ArXiv preprint: arXiv:0802.0886 (2008)

\bibitem{NielsenChuang}
Nielsen, M.A., Chuang, I.L.:
\newblock {Quantum Computation and Quantum Information}.
\newblock {Cambridge University Press} (October 2000)

\bibitem{OllingerJAC}
Ollinger, N.:
\newblock {Universalities in cellular automata a (short) survey.}
\newblock In Durand, B., ed.: First Symposium on Cellular Automata
  ``Journ{\'e}es Automates Cellulaires'' (JAC 2008), Uz{\`e}s, France, April
  21-25, 2008. Proceedings, MCCME Publishing House, Moscow (2008)  102--118

\bibitem{OllingerRichard}
Ollinger, N., Richard, G.:
\newblock {A Particular Universal Cellular Automaton.}
\newblock In Neary, T., Woods, D., Seda, A.K., Murphy, N., eds.: CSP, Cork
  University Press (2008)  267--278

\bibitem{PerezCheung}
P{\'e}rez-Delgado, C., Cheung, D.:
\newblock {Local unitary quantum cellular automata}.
\newblock Physical Review A \textbf{76}(3) (2007)  32320

\bibitem{Raussendorf}
Raussendorf, R.:
\newblock {Quantum cellular automaton for universal quantum computation}.
\newblock Phys. Rev. A \textbf{72}(022301) (2005)

\bibitem{SchumacherWerner}
Schumacher, B., Werner, R.:
\newblock {Reversible quantum cellular automata.}
\newblock ArXiv pre-print quant-ph/0405174 (2004)

\bibitem{ShepherdFranz}
Shepherd, D.J., Franz, T., Werner, R.F.:
\newblock {A universally programmable quantum cellular automata}.
\newblock Phys. Rev. Lett. \textbf{97}(020502) (2006)

\bibitem{Theyssier}
Theyssier, G.:
\newblock {Captive cellular automata}.
\newblock In: Proceedings of MFCS 2004, Lecture Notes in Computer Science,
  Springer (2004)  427--438

\bibitem{ToffoliConstruction}
Toffoli, T.:
\newblock {Computation and construction universality of reversible cellular
  automata}.
\newblock J. of Computer and System Sciences \textbf{15}(2) (1977)

\bibitem{VanDam}
Van~Dam, W.:
\newblock {Quantum cellular automata}.
\newblock Masters thesis, University of Nijmegen, The Netherlands (1996)

\bibitem{VollbrechtCirac}
Vollbrecht, K.G.H., Cirac, J.I.:
\newblock {Reversible universal quantum computation within
  translation-invariant systems}.
\newblock New J. Phys Rev A \textbf{73} (2004)  012324

\bibitem{Neumann}
von Neumann, J.:
\newblock {Theory of Self-Reproducing Automata}.
\newblock University of Illinois Press, Champaign, IL, USA (1966)

\bibitem{Watrous}
Watrous, J.:
\newblock {On one-dimensional quantum cellular automata}.
\newblock Complex Systems \textbf{5}(1) (1991)  19--–30

\bibitem{WatrousFOCS}
Watrous, J.:
\newblock {On one-dimensional quantum cellular automata}.
\newblock In: Proceedings of the 36th IEEE Symposium on Foundations of Computer
  Science, Washington, DC, USA, IEEE Computer Society (1995)  528--537

\end{thebibliography}

\bibliographystyle{splncs_srt}

\end{document}